\documentclass[preprint,showpacs,amsmath,amssymb,eqsecnum]{revtex4}
\usepackage{dcolumn}
\usepackage{bm}
\usepackage{graphicx}

\begin{document}
\title{{Can four-zero-texture mass matrix model reproduce the quark and lepton mixing angles and $CP$ violating phases?}}

\author{Koichi MATSUDA}
\affiliation{%
Center for High Energy Physics, 
Department of Engineering Physics,  
Tsinghua University, Beijing 100084, China}
\author{Hiroyuki NISHIURA}
\affiliation{%
Faculty of Information Science and Technology, 
Osaka Institute of Technology, 
Hirakata, Osaka 573-0196, Japan}

\date{June 12, 2006}

\begin{abstract}
We reconsider an universal mass matrix model which has a seesaw-invariant structure with four-zero texture common to all quarks and leptons.  
The CKM quark and MNS lepton mixing matrices of the model are analyzed analytically.
We show that the model can be consistent with all the experimental data of neutrino oscillation  
and quark mixings by tuning free parameters of the model. It is also shown that the model predicts a relatively large value  
for (1,3) element of the MNS lepton mixing matrix, $|(U_{MNS})_{13}|^2\simeq (0.041 - 9.6)\times 10^{-2}$. 
Using the seesaw mechanism, we also discuss the conditions for the components of the Dirac and the right-handed Majorana neutrino mass matrices 
which lead to the neutrino mass matrix consistent with the experimental data. 
\end{abstract}


\maketitle
\section{Introduction}
The discovery of neutrino oscillation~\cite{skamioka} indicates that 
neutrinos have finite masses and mix one another with near bimaximal lepton mixings 
in contrast to small quark mixings.  
In order to explain the large lepton mixings and small quark mixings, 
mass matrix models with various structures such as zero texture~\cite{fritzsch}--\cite{Kang}, 
flavor $2 \leftrightarrow 3$ symmetry~\cite{Fukuyama}-- \cite{Matsuda4} etc. 
have been investigated in the literature. 
We think that quarks and leptons should be unified. 
Therefore, it is an interesting approach to investigate a possibility that 
all the mass matrices of the quarks and leptons have the same form  
which can lead to the large lepton mixings and the small quark mixings simultaneously.  
Since the mass matrix model is intended to be embedded into a grand unified theory (GUT),  
it is desirable for the model to have the following features:  
(i) The structure is common to all the mass matrices,  
\(M_u\), \(M_d\), \(M_e\), and \(M_\nu \) for up quarks (\(u,c,t\)), down quarks (\(d,s,b\)),  
charged leptons (\(e,\mu,\tau \)), and  neutrinos (\(\nu_e,\nu_{\mu},\nu_{\tau} \)), respectively.  
(ii) Since we assume the seesaw mechanism~\cite{Yanagida} for neutrino masses,
the structure should conserve its form through the relation \(M_\nu \simeq -M_D M_R^{-1} M_D^T\).
We shall call this structure as a seesaw-invariant form. Here \(M_D\) and \(M_R\) are, respectively, the Dirac and the right-handed Majorana type neutrino mass matrices, 
which are also assumed to have the same structure.
\par
In this paper, as typical mass matrices which have the features mentioned above, 
we reconsider hermitian mass matrices $M_f$ for $f=u,d,e,\mbox{and }D$ and 
symmetric mass matrices $M_f$ for $f=\nu,\mbox{and }M$ with a universal form given by 
\begin{align}
M_f&=P^\dagger_f \widehat{M_f} P_f, \quad \mbox{for } f=u,d,e,\mbox{and }D,\\ 
M_f&=P^\dagger_f \widehat{M_f} P^{*}_f, \quad \mbox{for } f=\nu \mbox{ and }M .
\end{align}
Here $P_f$ is a diagonal phase matrix given by 
\begin{equation}
P_f =\mbox{diag}\left(e^{i\alpha_{f1}}, e^{i\alpha_{f2}}, e^{i\alpha_{f3}}\right),
\end{equation}
and the matrix $\widehat{M_f}$ is defined by
\begin{equation}
\widehat{M_f} \equiv \left(
\begin{array}{ccc}
0   & a_f & 0   \\
a_f & b_f & c_f\\
0   & c_f & d_f
\end{array}
\right),\label{M}
\end{equation}
for $f=u,d,e,\nu,D,\mbox{and }M$. 
In this seesaw-invariant type of four-zero-texture model, we have four real component parameters $a_f$, $b_f$, $c_f$, and $d_f$ 
in $\widehat{M_f}$ and phase parameters $\alpha_{fi}\, (i=1,2,3)$ in $P_f$.
If we fix three eigenvalues $m_{fi} \, (i=1,2, \mbox{and }3)$ of $\widehat{M_f}$ by the observed fermion masses, 
one free parameter is left in $\widehat{M_f}$. So we shall choose $d_f$ as the free parameter in this paper. 
Then we shall present analytical expressions for the Cabibbo-Kobayashi-Maskawa (CKM) quark mixing matrix~\cite{CKM} 
and the Maki-Nakagawa-Sakata (MNS) lepton mixing matrix~\cite{MNS} of the model in terms of $m_{f1}$, $m_{f2}$, $m_{f3}$, $d_f$ and $\alpha_{fi}$.
\par 
By taking a special value for this free parameter as $d_f=m_{f3}+m_{f1}$, 
the  model with the same structure has been discussed in Ref.~\cite{Nishiura2}.  
However, in this special choice, the model predicts a rather smaller value for (1,3) element of 
the CKM quark mixing matrix than the corresponding observed experimental data.
In order to overcome this defect in the quark sector, we treat $d_f$ as a free parameter in the present paper  
and show that the observed small CKM quark mixings as well as large MNS lepton mixings 
can be well derived by fine tuning of the free parameters.
\par
It has been claimed~\cite{Roberts, Kim, Branco} that four-zero-texture models for quarks 
are ruled out at the three $\sigma$ level from the experimental data for sin$2\beta$. 
However, we shall show from an analysis with use of the free parameter $d_f$ that the quark mixing angles and $CP$ violating phase $\delta_q$ in our model 
are consistent with the data at one $\sigma$ level, so that the sin$2\beta$ is also consistent at the same level.  
\par
This article is organized as follows. 
In Sec.~II, we discuss the diagonalization of mass matrix of our model. 
In Sec.~III, approximations we use are presented. 
The analytical expressions of the quark mixing matrix of the model are given in Sec.~IV. 
In Sec.~V, the lepton mixing matrix of the model is given.
Sec.~VI  is devoted to a summary.

\section{Diagonalization of Mass matrix}
\par
We now discuss a diagonalization of the mass matrix $M_f$. 
First we argue a diagonalization of $\widehat{M_f}$ given by 
\begin{equation}
\widehat{M_f}=
\left(
\begin{array}{lll}
\ 0 & \ a_f & \ 0 \\
\ a_f & \ b_f & \ c_f \\
\ 0 & \ c_f & \ d_f \\
\end{array}
\right).\label{M_hat_f}
\end{equation}
This is diagonalized  by an orthogonal matrix $O_f$ as 
\begin{equation}
O_f^T \widehat{M_f} O_f = \mbox{diag}(m_{f1} , m_{f2} , m_{f3}),
\end{equation}
where $m_{f1}$, $m_{f2}$, and $m_{f3}$ are eigenvalues of $M_f$.
Here we have four component parameters in $\widehat{M_f}$, namely, $a_f$, $b_f$, $c_f$, and $d_f$.
If we fix the $m_{fi}$ by the observed quark and/or lepton mass, 
we have one free parameter left. Therefore we choose $d_f$ as the free parameter. 
Then, we derive explicit expressions of the orthogonal matrix $O_f$ 
in terms of $m_{f1}$, $m_{f2}$, $m_{f3}$, and $d_f$ as
\begin{equation}
O_f = \left(
\begin{array}{ccc}
 \sqrt{ \frac{(d_f-m_{f1})m_{f2}m_{f3}      }{R_{f1}d_f}} &
 \sqrt{ \frac{(d_f-m_{f2})m_{f3}m_{f1}      }{R_{f2}d_f}} &
 \sqrt{ \frac{(d_f-m_{f3})m_{f1}m_{f2}      }{R_{f3}d_f}} \\
-\sqrt{-\frac{(d_f-m_{f1})m_{f1}         }{R_{f1}   }} &
 \sqrt{-\frac{(d_f-m_{f2})m_{f2}         }{R_{f2}   }} &
 \sqrt{-\frac{(d_f-m_{f3})m_{f3}         }{R_{f3}   }} \\
 \sqrt{ \frac{m_{f1}(d_f-m_{f2})(d_f-m_{f3})}{R_{f1}d_f}} &
-\sqrt{ \frac{m_{f2}(d_f-m_{f3})(d_f-m_{f1})}{R_{f2}d_f}} &
 \sqrt{ \frac{m_{f3}(d_f-m_{f1})(d_f-m_{f2})}{R_{f3}d_f}}
\end{array}
\right),\label{O_f}
\end{equation}
where $R_{fi}\, (i=1,2,\mbox{and } 3)$ are defined by
\begin{align}
R_{f1} &= (m_{f1}-m_{f2})(m_{f1}-m_{f3}) , \\
R_{f2} &= (m_{f2}-m_{f3})(m_{f2}-m_{f1}) , \\
R_{f3} &= (m_{f3}-m_{f1})(m_{f3}-m_{f2}).
\end{align}
The expressions of the components $a_f$, $b_f$, and $c_f$
in terms of $m_{f1}$, $m_{f2}$, $m_{f3}$, and $d_f$ are presented as 
\begin{align}
a_f &= \sqrt{-\frac{m_{f1}m_{f2}m_{f3}}{d_f}}, \\
b_f &= m_{f1} + m_{f2} + m_{f3} - d_f, \\
c_f &= \sqrt{-\frac{(d_f-m_{f1})(d_f-m_{f2})(d_f-m_{f3})}{d_f}}. 
\end{align}
From the condition that $a_f$, $b_f$, and $c_f$ are real, we have the allowed region of $d_f$ given by
\begin{equation}
|m_{f1}| <  d_f < |m_{f3}| .
\end{equation}
The cases in which $0 <  d_f < |m_{f1}|$ or $|m_{f3}| <  d_f $ are not allowed.
We also have the following sign assignments for the eigenmass $m_{fi}$:
\begin{eqnarray}
 0 <  m_{f1} < -m_{f2} < m_{f3}  & &\mbox{for } \quad    |m_{f1}| <  d_f < |m_{f2}|,\\ 
 0 < -m_{f1} <  m_{f2} < m_{f3}  & &\mbox{for } \quad    |m_{f2}| <  d_f < |m_{f3}|.
\end{eqnarray}
Namely $m_{f2}$  should be taken negative while $m_{f1}$ and $m_{f3}$ are positive for the case in which $|m_{f1}| <  d_f < |m_{f2}|$. 
On the other hand, $m_{f1}$ should be taken negative while $m_{f2}$ and $m_{f3}$ are positive for $|m_{f2}| <  d_f < |m_{f3}|$.

\section{approximations}
\par
We present approximated expressions of the orthogonal matrix $O_f$ for the normal hierarchy, inverse hierarchy, 
and quasi degenerate cases for the masses $m_{fi}$. Here we introduce a $x_f$ parameter, instead of using $d_f$, defined by 
\begin{equation}
x_f = \frac{d_f}{m_{f3}}.
\end{equation}
The approximated expressions are obtained as follows:
\par
Case (a): For \(|m_{f1}| \ll m_{f2} \ll  d_f < m_{f3} \) (normal hierarchy 1), we have 
\begin{equation}
O_f \simeq \left(
\begin{array}{ccc}
1 & 
\sqrt{\frac{|m_{f1}|}{m_{f2}}} &  
\sqrt{\frac{|m_{f1}|m_{f2}}{m_{f3}^2} \frac{1-x_f}{x_f}} \\
-\sqrt{\frac{|m_{f1}|}{m_{f2}} x_f} &
 \sqrt{x_f} &
 \sqrt{1-x_f} \\
 \sqrt{\frac{|m_{f1}|}{m_{f2}} (1-x_f)} &
 -\sqrt{1-x_f} &
 \sqrt{x_f}
\end{array}
\right).\label{O_f_a}
\end{equation}

Case (b): For \(|m_{f1}| < m_{f2} \ll  d_f < m_{f3} \) (normal hierarchy 2), we have
\begin{equation}
O_f \simeq \left(
\begin{array}{ccc}
\sqrt{\frac{|m_{f2}|}{|m_{f1}|+m_{f2}}} & 
\sqrt{\frac{|m_{f1}|}{|m_{f1}|+m_{f2}}} &  
\sqrt{\frac{|m_{f1}|m_{f2}}{m_{f3}^2} \frac{1-x_f}{x_f}} \\
-\sqrt{\frac{|m_{f1}|}{|m_{f1}|+m_{f2}} x_f} &
 \sqrt{\frac{ m_{f2} }{|m_{f1}|+m_{f2}} x_f} &
 \sqrt{1-x_f} \\
 \sqrt{\frac{|m_{f1}|}{|m_{f1}|+m_{f2}} (1-x_f)} &
 -\sqrt{\frac{ m_{f2} }{|m_{f1}|+m_{f2}} (1-x_f)} &
 \sqrt{x_f}
\end{array}
\right).\label{O_f_b}
\end{equation}

Case (c): For \(m_{f1} \ll d_f < |m_{f2}| \simeq m_{f3} \) (inverse hierarchy), we have
\begin{equation}
O_f \simeq \left(
\begin{array}{ccc}
1 & 
\sqrt{\frac{m_{f3} m_{f1} (d_f+|m_{f2}|)}{|m_2|(|m_{f2}|+m_{f3})d_f}} &  
\sqrt{\frac{m_{f1}|m_{f2}|(d_f+|m_{f2}|)}{|m_3|(|m_{f2}|+m_{f3})d_f}} \\
-\sqrt{\frac{ m_{f1} d_f}{|m_{f2}|m_{f3}}} &
 \sqrt{\frac{ d_f + |m_{f2}| }{|m_{f2}|+m_{f3}}} &
 \sqrt{\frac{ m_{f3} - d_f }{|m_{f2}|+m_{f3}}} \\
 \sqrt{\frac{m_{f1} (d_f+|m_{f2}|)(m_{f3} - d_f)}{|m_2| m_{f3} d_f}} &  
-\sqrt{\frac{ m_{f3} - d_f }{|m_{f2}|+m_{f3}}} &
 \sqrt{\frac{ d_f + |m_{f2}| }{|m_{f2}|+m_{f3}}}
\end{array}
\right).\label{O_f_c}
\end{equation}

Case (d): For \(m_f = |m_{f1}| < m_{f2} < d_f <  m_{f3} \) (quasi degenerate 1), we have
\begin{equation}
O_f \simeq \left(
\begin{array}{ccc}
 \sqrt{\frac{1}{2}} & 
 \sqrt{\frac{1}{2} \frac{d_f - m_{f2}}{m_{f3}-m_{f2}}} &  
 \sqrt{\frac{1}{2} \frac{m_{f3} - d_f}{m_{f3}-m_{f2}}}  \\
-\sqrt{\frac{1}{2}} & 
 \sqrt{\frac{1}{2} \frac{d_f - m_{f2}}{m_{f3}-m_{f2}}} &  
 \sqrt{\frac{1}{2} \frac{m_{f3} - d_f}{m_{f3}-m_{f2}}}  \\
 \sqrt{\frac{1}{4} \frac{(d_f - m_{f2})(m_{f3} - d_f)}{m_f^2}} & 
-\sqrt{\frac{m_{f2} - d_f}{m_{f3}-m_{f2}}} &
 \sqrt{\frac{d_f - m_{f2}}{m_{f3}-m_{f2}}}  
\end{array}
\right).\label{O_f_d}
\end{equation}

Case (e): For \(m_f = m_{f1} < d_f < |m_{f2}| < m_{f3} \) (quasi degenerate 2), we have
\begin{equation}
O_f \simeq \left(
\begin{array}{ccc}
 \sqrt{\frac{1}{2} \frac{d_f - m_{f1}}{m_{f3}-m_{f1}}} &  
 \sqrt{\frac{1}{2}} & 
 \sqrt{\frac{1}{2} \frac{m_{f3} - d_f}{m_{f3}-m_{f1}}}  \\
-\sqrt{\frac{1}{2} \frac{d_f - m_{f1}}{m_{f3}-m_{f1}}} &  
 \sqrt{\frac{1}{2}} & 
 \sqrt{\frac{1}{2} \frac{m_{f3} - d_f}{m_{f3}-m_{f1}}}  \\
 \sqrt{\frac{m_{f3} - d_f}{m_{f3}-m_{f1}}} &
-\sqrt{\frac{1}{4} \frac{(m_{f3} - d_f)(d_f - m_{f1})}{m_f^2}} & 
 \sqrt{\frac{d_f - m_{f1}}{m_{f3}-m_{f1}}}  
\end{array}
\right).\label{O_f_e}
\end{equation}
The inverse hierarchy and the quasi degenerate scenarios are
unfavorable in our model.
\section{CKM quark mixing matrix}
Let us discuss the quark sector.
The mass matrices $M_u$ and $M_d$ for the u- and d-quarks are, respectively, given by
\begin{eqnarray}
M_u&=&P^\dagger_u \widehat{M_u}P_u, \\
M_d&=&P^\dagger_d \widehat{M_d}P_d,
\end{eqnarray}
where $P_{u}$ and $P_{d}$ are diagonal phase matrices and  
$\widehat{M_u}$ and $\widehat{M_d}$ are given by Eq.~(\ref{M}). 
The mass matrix $M_f$ ($f=u\mbox{ and }d$) are diagonalized as 
\begin{equation}
U_{Lf}^\dagger M_f U_{Lf} = \mbox{diag}\left(-|m_{f1}|, m_{f2}, m_{f3}\right).
\end{equation}
The unitary matrix $U_{Lf}$ is described as
\begin{equation}
U_{Lf} =P_{f}^\dagger O_f.
\end{equation}
Therefore the CKM quark mixing matrix $U_{CKM}$ 
of the model is given by 
\begin{equation}
U_{CKM}=U^\dagger_{Lu}U_{Ld}=O^{T}_u PO_d,\label{our_ckm}
\end{equation}
where $P \equiv P_uP^\dagger_d$ is diagonal phase matrix given by
\begin{equation}
P 
=\mbox{diag}(e^{i(\alpha_{u1}-\alpha_{d1})} , e^{i(\alpha_{u2}-\alpha_{d2})} , e^{i(\alpha_{u3}-\alpha_{d3})})
\equiv \mbox{diag}(1 , e^{i\alpha_2} , e^{i\alpha_3}).
\end{equation}
Here we take $\alpha_{d1}=\alpha_{u1}=0$ without any loss of generality.
\par
By using the expressions of $O_d$ and $O_u$ in Eq.~(\ref{O_f}), the explicit $(i,j)$ elements of $U_{CKM}$ 
are obtained as
\begin{align}
(U_{CKM})_{12} =&
 \sqrt{ \frac{m_c m_t (d_u-m_u)}{R_{u1} d_u}} 
 \sqrt{ \frac{m_b m_d (d_d-m_s)}{R_{d2} d_d}}
 -e^{i \alpha_2}
 \sqrt{-\frac{m_u (d_u-m_u)}{R_{u1}}} 
 \sqrt{-\frac{m_s (d_d-m_s)}{R_{d2}}} \nonumber \\
&-e^{i \alpha_3}
 \sqrt{ \frac{m_u (d_u-m_c) (d_u-m_t)}{R_{u1} d_u}} 
 \sqrt{ \frac{m_s (d_d-m_b) (d_d-m_d)}{R_{d2} d_d}} \label{exact_our_ckm_1} \\
(U_{CKM})_{13} =&
 \sqrt{ \frac{m_c m_t (d_u-m_u)}{R_{u1} d_u}} 
 \sqrt{ \frac{m_d m_s (d_d-m_b)}{R_{d3} d_d}}
 -e^{i \alpha_2}
 \sqrt{-\frac{m_u (d_u-m_u)}{R_{u1}}} 
 \sqrt{-\frac{m_b (d_d-m_b)}{R_{d3}}} \nonumber \\
&-e^{i \alpha_3}
 \sqrt{ \frac{m_u (d_u-m_c) (d_u-m_t)}{R_{u1} d_u}} 
 \sqrt{ \frac{m_b (d_d-m_d) (d_d-m_s)}{R_{d3} d_d}}\label{exact_our_ckm_2} \\
(U_{CKM})_{23} =&
 \sqrt{ \frac{m_t m_u (d_u-m_c)}{R_{u2} d_u}} 
 \sqrt{ \frac{m_d m_s (d_d-m_b)}{R_{d3} d_d}}
 -e^{i \alpha_2}
 \sqrt{-\frac{m_c (d_u-m_c)}{R_{u2}}} 
 \sqrt{-\frac{m_b (d_d-m_b)}{R_{d3}}} \nonumber \\
&-e^{i \alpha_3}
 \sqrt{ \frac{m_c (d_u-m_t) (d_u-m_u)}{R_{u2} d_u}} 
 \sqrt{ \frac{m_b (d_d-m_d) (d_d-m_s)}{R_{d3} d_d}}\label{exact_our_ckm_3} 
\end{align}
where $R_{ui}$ and $R_{di}\, (i=1,2,\mbox{and } 3)$ are given  by
\begin{align}
R_{u1} &= (m_{u}-m_{c})(m_{u}-m_{t}) , \\
R_{u2} &= (m_{c}-m_{t})(m_{c}-m_{u}) , \\
R_{u3} &= (m_{t}-m_{u})(m_{t}-m_{c}).
\end{align}
\begin{align}
R_{d1} &= (m_{d}-m_{s})(m_{d}-m_{b}) , \\
R_{d2} &= (m_{s}-m_{b})(m_{s}-m_{d}) , \\
R_{d3} &= (m_{b}-m_{d})(m_{b}-m_{s}).
\end{align}
Here, we denoted $m_{ui}$ and $m_{di}\ (i=1,2,3)$ as $(m_u, m_c, m_t)$ and $(m_d, m_s, m_b)$  
which are the masses of up and down quarks, respectively. 
\par 
If we fix the quark masses $(m_u, m_c, m_t)$ and $(m_d, m_s, m_b)$ by the observed masses, 
two component parameters $d_u$ and $d_d$ and two phase parameters $\alpha_2$ and $\alpha_3$ are left as free parameters in above expressions of $(U_{CKM})_{ij}$. 
Using this feature of the model, we can reproduce the observed data for $(U_{CKM})_{ij}$  as will be shown later. 
This model can be used for the improvement of the previous model~\cite{Nishiura2} in which a rather small value for $|(U_{CKM})_{13}|$ is predicted. 

In the discussions of the CKM quark mixing matrix, we concentrate our attention on the case in which
\(|m_{f1}| \ll m_{f2} \ll d_f < m_{f3}\) (normal hierarchy 1) .
In this case, using two free parameters \(x_u \equiv d_u/m_{t}\) and \(x_d \equiv d_d/m_{b}\) instead of using $d_u$ and $d_d$, we have
\begin{align} 
&(U_{CKM})_{12} \simeq
                 \sqrt{\frac{|m_d|}{m_s}}
 -e^{i \alpha_2} \sqrt{\frac{|m_u|}{m_c} x_u x_d}
 -e^{i \alpha_3} \sqrt{\frac{|m_u|}{m_c} (1-x_u) (1- x_d)}\, , \label{our_ckm_1}\\
&(U_{CKM})_{13} \simeq
                 \sqrt{\frac{|m_d| m_s}{m_b^2} \frac{1-x_d}{x_d}}
 -e^{i \alpha_2} \sqrt{\frac{|m_u|}{m_c} x_u (1-x_d)}
 +e^{i \alpha_3} \sqrt{\frac{|m_u|}{m_c} (1-x_u) x_d}\, , \label{our_ckm_2}\\
&(U_{CKM})_{23} \simeq
                 \sqrt{\frac{|m_u|}{m_c} \frac{|m_d| m_s}{m_b^2} \frac{1-x_d}{x_d}}
 +e^{i \alpha_2} \sqrt{x_u (1-x_d)}
 -e^{i \alpha_3} \sqrt{(1-x_u) x_d}\, .\label{our_ckm_3}
\end{align}
\par
By using the rephasing of the up and down quarks, 
Eq.~(\ref{our_ckm}) is changed to the standard representation of the CKM quark mixing matrix, 
\begin{eqnarray}
U_{CKM}^{\rm std} &=& \mbox{diag}(e^{i\zeta_1^u},e^{i\zeta_2^u},e^{i\zeta_2^u})  \ U_{CKM} \ 
\mbox{diag}(e^{i\zeta_1^d},e^{i\zeta_2^d},e^{i\zeta_2^d}) \nonumber \\
&=&
\left(
\begin{array}{ccc}
c_{13}^qc_{12}^q & c_{13}^qs_{12}^q & s_{13}^qe^{-i\delta_q} \\
-c_{23}^qs_{12}^q-s_{23}^qc_{12}^qs_{13}^q e^{i\delta_q}
&c_{23}^qc_{12}^q-s_{23}^qs_{12}^qs_{13}^q e^{i\delta_q} 
&s_{23}^qc_{13}^q \\
s_{23}^qs_{12}^q-c_{23}^qc_{12}^qs_{13}^q e^{i\delta_q}
 & -s_{23}^qc_{12}^q-c_{23}^qs_{12}^qs_{13}^q e^{i\delta_q} 
& c_{23}^qc_{13}^q \\
\end{array}
\right) \ .
\label{stdrep}
\end{eqnarray}
Here \(\zeta_i^q\) comes from the rephasing in the quark fields 
to make the choice of phase convention.
By using the expressions of $U_{CKM}$ in Eqs.~(\ref{our_ckm_1})-(\ref{our_ckm_3}), 
the $CP$ violating phase $\delta_q$ in the quark mixing matrix is given by
\begin{eqnarray}
\delta_q &=&
\mbox{arg}\left[
\left(\frac{(U_{CKM})_{12} (U_{CKM})_{22}^{*}}{(U_{CKM})_{13} (U_{CKM})_{23}^{*}}\right) + 
\frac{|(U_{CKM})_{12}|^2}{1-|(U_{CKM})_{13}|^2}
\right] \label{exact_our_delta_q} \\
&\simeq &
\mbox{arg} \Biggl[ 
   \frac{\left(e^{i \alpha_3} \sqrt{(1-x_u)(1-x_d)} + e^{i \alpha_2} \sqrt{x_u x_d}\right)^*}
        {\left(
          e^{i \alpha_3} \sqrt{(1-x_u) x_d} - e^{i \alpha_2} \sqrt{x_u (1-x_d)}
         \right)
         \left(
          e^{i \alpha_2} \sqrt{x_u(1-x_d)} - e^{i \alpha_3} \sqrt{(1-x_u) x_d}
         \right)^*}
        \Biggr]. \nonumber \\
\label{our_delta_q}
\end{eqnarray}
\par
Thus we have obtained the analytical expressions for $|(U_{CKM})_{12}|$, $|(U_{CKM})_{23}|$, $|(U_{CKM})_{13}|$, and $\delta_q$ of the model 
which are given by Eqs.~(\ref{our_ckm_1}), (\ref{our_ckm_2}), (\ref{our_ckm_3}), and (\ref{our_delta_q}), respectively.
They are functions of the four parameters $x_u$, $x_d$, $\alpha_2$, and $\alpha_3$. 
From the expressions of $|(U_{CKM})_{13}|$ and $|(U_{CKM})_{23}|$ 
in Eqs.~(\ref{our_ckm_2}) and (\ref{our_ckm_3}),
we obtain the following constraints in the paprameters $x_u$ and $x_d$, 
which hold irrespectively of the free phase parameters $\alpha_2$ and $\alpha_3$.
\begin{eqnarray}
&&\frac{1}{1+\frac{\sqrt{\frac{|m_u|}{m_c}}|(U_{CKM})_{23}|+|(U_{CKM})_{13}|}{\frac{|m_d|m_s}{m_b^2} }} 
\alt x_d \alt \frac{1}{1+\frac{\sqrt{\frac{|m_u|}{m_c}}|(U_{CKM})_{23}|-|(U_{CKM})_{13}|}{\frac{|m_d|m_s}{m_b^2} }}, \\
&& \left|\sqrt{x_u(x_d-1)}-\sqrt{x_d(x_u-1)}\right| \alt |(U_{CKM})_{23}| \alt \left|\sqrt{x_u(x_d-1)}+\sqrt{x_d(x_u-1)}\right|.
\end{eqnarray}
\par
On the other hand, the numerical values of $|(U_{CKM})_{12}|$, $|(U_{CKM})_{23}|$, $|(U_{CKM})_{13}|$, and $\delta_q$ 
at the unification scale \(\mu=M_X\)  are estimated from the experimental data observed at electroweak scale \(\mu=M_Z\) 
by using the renormalization group equation as~\cite{Matsuda3}:
\begin{eqnarray}
|(U_{CKM})_{12}|&=&0.2226-0.2259,\label{U_CKM_BD_12}\\
|(U_{CKM})_{23}|&=&0.0295-0.0387,\label{U_CKM_BD_23}\\
|(U_{CKM})_{13}|&=&0.0024-0.0038,\label{U_CKM_BD_13}\\
\delta_q &=& 46^\circ   - 74^\circ . 
\end{eqnarray}
\par
By using the above experimental constraints as inputs, 
we obtain the consistent solution for the parameter $x_u$, $x_d$, $\alpha_2$, and $\alpha_3$ of our model 
from our exact CKM matrix elements given by Eqs.~(\ref{exact_our_ckm_1}), (\ref{exact_our_ckm_2}), (\ref{exact_our_ckm_3}), and (\ref{exact_our_delta_q}). 
By doing parameter fitting, we find that the consistent CKM elements are realized only if 
(i) the parameter $\alpha_2$ takes a value as 
$\alpha_2 \simeq \pi/2$ 
and (ii) the other three parameters $\alpha_3$, $x_u$, and $x_d$ take values in the allowed regions shown in Fig.~1, Fig.~2, and Fig.~3.
The best fit is realized for the following values of the parameters:
\begin{eqnarray}
\alpha_2 &=& \pi/2,\\
\alpha_3 &=& 1.450,\\
x_u &=& 0.9560,\\
x_d &=& 0.9477.
\end{eqnarray}
For these best-fit-parameters of the model, we obtain
\begin{eqnarray}
|(U_{CKM})_{12}|&=&0.2251,\\
|(U_{CKM})_{23}|&=&0.0340,\\
|(U_{CKM})_{13}|&=&0.0032,\\
\delta_q &=& 58.86^\circ . 
\end{eqnarray}
Here we have used the best fit values of the following quark masses estimated~\cite{Fusaoka} at the unification scale \(\mu=M_X\).
\begin{equation}
\begin{array}{ll}
m_u(M_X)=1.04^{+0.19}_{-0.20}\, \mbox{MeV},& 
m_d(M_X)=1.33^{+0.17}_{-0.19}\, \mbox{MeV}, \\
m_c(M_X)=302^{+25}_{-27}\, \mbox{MeV}, &
m_s(M_X)=26.5^{+3.3}_{-3.7}\, \mbox{MeV}, \\ 
m_t(M_X)=129^{+196}_{-40}\,  \mbox{GeV}, &
m_b(M_X)=1.00\pm0.04\, \mbox{GeV}. 
\end{array}
\label{eq123104}
\end{equation}
\par
Finally let us mention the model in Ref.~\cite{Nishiura2}. 
It corresponds to our present model with  
the parameter \(d_f\) fixed as \(d_f=m_{f3}+m_{f1}\), namely  \(x_f = 1 -|m_{f1}|/m_{f3} \simeq 1\) 
and \(1-x_f= |m_{f1}|/m_{f3}\).  
In this case,  the following CKM mixing matrix elements are derived 
as seen from Eqs.~(\ref{our_ckm_1})-(\ref{our_ckm_3}) and (\ref{our_delta_q}).
\begin{align} 
&(U_{CKM})_{12} \simeq
                 \sqrt{\frac{|m_d|}{m_s}}
 -e^{i \alpha_2} \sqrt{\frac{|m_u|}{m_c}}
 -e^{i \alpha_3} \sqrt{\frac{|m_u^2|}{m_c m_t} \frac{|m_d|}{m_b}},\\
&(U_{CKM})_{13} \simeq
                 \sqrt{\frac{|m_d^2| m_s}{m_b^3} }
 -e^{i \alpha_2} \sqrt{\frac{|m_u|}{m_c} \frac{|m_d|}{m_b}}
 +e^{i \alpha_3} \sqrt{\frac{|m_u|^2}{m_c m_t}}, \\
&(U_{CKM})_{23} \simeq
                 \sqrt{\frac{|m_u|}{m_c} \frac{|m_d^2| m_s}{m_b^3}}
 +e^{i \alpha_2} \sqrt{\frac{|m_d|}{m_b}}
 -e^{i \alpha_3} \sqrt{\frac{|m_u|}{m_t}}, \\
&\delta_q  \simeq \mbox{arg} \Biggl[ 
   \frac{\left(e^{i \alpha_3} \sqrt{\frac{|m_u|}{m_t}\frac{|m_d|}{m_b}} + e^{i \alpha_2}\right)^* }
        {\left(
          e^{i \alpha_3} \sqrt{\frac{|m_u|}{m_t}}
           - e^{i \alpha_2} \sqrt{\frac{|m_d|}{m_b}}
         \right)
         \left(
          e^{i \alpha_2} \sqrt{\frac{|m_d|}{m_b}}
           - e^{i \alpha_3} \sqrt{\frac{|m_u|}{m_t}}
         \right)^*}
        \Biggr]\simeq \pi-\alpha_2.
\end{align}
This model is more predictable for the CKM matrix elements than ours. 
However, this model predicts a rather smaller value for $|(U_{CKM})_{13}|$ than the experimental data. 
This is seen from the fact that the values of the parameters $x_d=1 -|m_{d}|/m_{b}=0.998437$ and $x_u=1 -|m_{u}|/m_{t}=0.999987$ of this model are 
outside of the allowed regions shown in Fig.~1, Fig.~2, and Fig.~3.  

\section{MNS lepton mixing matrix}
Let us discuss the lepton sector.
The mass matrices $M_\nu$ and $M_e$ for the Majorana neutrinos and the charged leptons are, respectively given by
\begin{eqnarray}
M_e&=&P^\dagger_e \widehat{M_e}P_e,\\
M_\nu&=&P^\dagger_\nu \widehat{M_\nu}P^{*}_\nu. 
\end{eqnarray}
Here $P_{\nu}$ and $P_{e}$ are diagonal phase matrices and  
$\widehat{M_\nu}$ and $\widehat{M_e}$ are given by Eq.~(\ref{M_hat_f}). 
The charged lepton mass matrix $M_e$ is diagonalized as 
\begin{equation}
U_{Le}^\dagger M_e U_{Le} = \mbox{diag}\left(-|m_e|, m_{\mu}, m_{\tau}\right),
\end{equation}
where the unitary matrix $U_{e}$ is described as
\begin{equation}
U_{Le} =P_{e}^\dagger O_e.
\end{equation}
Since the mass matrix for the Majorana neutrinos is symmetric, 
$M_\nu$ is diagonalized as 
\begin{equation}
U_{\nu}^\dagger M_\nu U_{\nu}^* = \mbox{diag}\left(|m_{1}|, m_{2}, m_{3}\right),
\end{equation}
where $|m_{1}|, m_{2}, \mbox{and } m_{3}$ are real positive neutrino masses and  
the unitary matrix $U_{\nu}$ is described as
\begin{equation}
U_{\nu} =P_{\nu}^\dagger O_\nu Q_{\nu}.
\end{equation}
Here, in order to make the neutrino masses for the first generation to be real positive, 
we introduce an additional diagonal phase matrix $Q_{\nu}$ defined by 
\begin{equation}
Q_{\nu} \equiv \mbox{diag}\left(i,1,1\right). \ \label{Q}
\end{equation}
\par
In the following discussions, we consider the normal hierarchy 2 
for the neutrino masses $m_i$, 
i.e. $|m_1|< m_2 \ll d_\nu <m_3$, and the normal hierarchy 1 
for the charged lepton masses,  
i.e. $|m_e| \ll m_\mu \ll d_e <m_\tau$. 
In this case, the orthogonal matrix $O_e$ and $O_\nu$ are obtained 
from Eqs.~(\ref{O_f_a}) and (\ref{O_f_b})  with $f=e\,\,\mbox{and }\nu$ by replacing \(|m_{1f}|\), \(m_{2f}\), and \(m_{3f}\) 
with \(|m_{e}|\), \(m_{\mu}\), and \(m_{\tau}\), and with \(|m_{1}|\), \(m_{2}\), and \(m_{3}\), respectively.
Therefore we have 
\begin{eqnarray} 
O_\nu &\simeq& \left(
\begin{array}{ccc}
\sqrt{\frac{m_{2}}{|m_{1}|+m_{2}}} & 
\sqrt{\frac{|m_{1}|}{|m_{1}|+m_{2}}} &  
\sqrt{\frac{|m_{1}|m_{2}}{m_{3}^2} \frac{1-x_\nu}{x_\nu}} \\
-\sqrt{\frac{|m_{1}|}{|m_{1}|+m_{2}} x_\nu} &
 \sqrt{\frac{m_{2}}{|m_{1}|+m_{2}}x_\nu} &
 \sqrt{1-x_\nu} \\
 \sqrt{\frac{|m_{1}|}{|m_{1}|+m_{2}} (1-x_\nu)} &
 -\sqrt{\frac{m_{2}}{|m_{1}|+m_{2}}(1-x_\nu)} &
 \sqrt{x_\nu}
\end{array}
\right), \\
O_e &\simeq& \left(
\begin{array}{ccc}
1 & 
\sqrt{\frac{|m_{e}|}{m_{\mu}}} &  
\sqrt{\frac{|m_{e}|m_{\mu}}{m_{\tau}^2} \frac{1-x_e}{x_e}} \\
-\sqrt{\frac{|m_{e}|}{m_{\mu}} x_e} &
 \sqrt{x_e} &
 \sqrt{1-x_e} \\
 \sqrt{\frac{|m_{e}|}{m_{\mu}} (1-x_e)} &
 -\sqrt{1-x_e} &
 \sqrt{x_e}
\end{array}
\right).
\end{eqnarray}
We now discuss the MNS lepton mixing matrix $U_{MNS}$ 
of the model, which is given by 
\begin{equation}
U_{MNS}=U^\dagger_{Le}U_{\nu}=O^{T}_eP_{\ell}O_\nu Q_\nu,
\end{equation}
where $P_{\ell} \equiv P_eP^\dagger_\nu$ is diagonal phase matrix and we take 
\begin{equation}
P_{\ell} =\mbox{diag}(1 , e^{i\beta_2} , e^{i\beta_3}),
\end{equation}
without any loss of generality.
Thus we obtain 
\begin{equation}
U_{MNS} \simeq \left(
\begin{array}{ccc}
i\sqrt{\frac{m_{2}}{|m_{1}|+m_{2}}} & 
\sqrt{\frac{|m_{1}|}{|m_{1}|+m_{2}}} &  
\sqrt{\frac{|m_{1}|m_{2}}{m_{3}^2} \frac{1-x_\nu}{x_\nu}} +\xi_5\sqrt{\frac{|m_e|}{m_\mu}}\\
-i\xi_1\sqrt{\frac{|m_{1}|}{|m_{1}|+m_{2}}}  &
 \xi_1\sqrt{\frac{m_{2}}{|m_{1}|+m_{2}}} &
 \xi_2 \\
 i\xi_3\sqrt{\frac{|m_{1}|}{|m_{1}|+m_{2}}} &
 -\xi_3\sqrt{\frac{m_{2}}{|m_{1}|+m_{2}}} &
 \xi_4
\end{array}
\right), 
\label{eq06061201}
\end{equation}
where $\xi_i$ are complex quantities defined by
\begin{eqnarray}
\xi_1&=&\sqrt{x_\nu x_e}e^{i \beta_2}+\sqrt{(1-x_\nu)(1- x_e)}e^{i \beta_3},\\
\xi_2&=&\sqrt{(1-x_\nu) x_e}e^{i \beta_2}-\sqrt{x_\nu(1- x_e)}e^{i \beta_3},\\
\xi_3&=&-\sqrt{x_\nu (1-x_e)}e^{i \beta_2}+\sqrt{(1-x_\nu)x_e}e^{i \beta_3},\\
\xi_4&=&\sqrt{(1-x_\nu)(1- x_e)}e^{i \beta_2}+\sqrt{x_\nu x_e}e^{i \beta_3},\\
\xi_5&=&\sqrt{(1-x_\nu) x_e}e^{i \beta_2}+\sqrt{x_\nu(1- x_e)}e^{i \beta_3}.
\end{eqnarray}
Eq.~(\ref{eq06061201}) is changed to the standard representation of the MNS 
lepton mixing matrix as well as the CKM quark mixing matrix, 
\begin{eqnarray}
U_{MNS}^{\rm std} &=& \mbox{diag}(e^{i\zeta_1^e},e^{i\zeta_2^e},e^{i\zeta_2^e})  
\ U_{MNS}  \nonumber \\
&=&
\left(
\begin{array}{ccc}
c_{13}^lc_{12}^l & c_{13}^ls_{12}^l & s_{13}^le^{-i\delta_l} \\
-c_{23}^ls_{12}^l-s_{23}^lc_{12}^ls_{13}^l e^{i\delta_l}
&c_{23}^lc_{12}^l-s_{23}^ls_{12}^ls_{13}^l e^{i\delta_l} 
&s_{23}^lc_{13}^l \\
s_{23}^ls_{12}^l-c_{23}^lc_{12}^ls_{13}^l e^{i\delta_l}
 & -s_{23}^lc_{12}^l-c_{23}^ls_{12}^ls_{13}^l e^{i\delta_l} 
& c_{23}^lc_{13}^l \\
\end{array}
\right) \nonumber \\
&&\times \mbox{diag}(1,e^{i \phi_2},e^{i \phi_3}) .
\label{stdrepLep}
\end{eqnarray}
Here \(\zeta_i^e\) comes from the rephasing in the charged-lepton fields,
\(\delta_\nu\) is the Dirac phase,
and \(\phi_i\) is the Majorana phases in the MNS lepton mixing matrix.
\par
In order to realize the maximal lepton  mixing angle between the second and third generations, 
we must choose the free parameters $x_\nu$, $x_e$, $\beta_2$ and $\beta_3$ 
to satisfy the following condition:
\begin{equation}
|\xi_1|=|\xi_2|=|\xi_3|=|\xi_4|=\sqrt{\frac{1}{2}}.
\end{equation}
In the present paper, we take the following choice:
\begin{equation}
x_{\nu}=1/2 \quad \mbox{and } x_e \simeq 1,
\end{equation}
which satisfies the above condition irrespectively of the phases $\beta_2$ and  $\beta_3$.
Then, the explicit magnitudes of the components of $|(U_{MNS})_{ij}|$ are obtained as
\begin{align}
\left|(U_{MNS})_{11}\right|  & \simeq \sqrt{\frac{m_2}{m_2+|m_1|}}, & 
\left|(U_{MNS})_{12}\right|  & \simeq \sqrt{\frac{|m_1|}{m_2+|m_1|}}, \nonumber \\
\left|(U_{MNS})_{13}\right|  & \simeq \left|\sqrt{\frac{|m_1|m_2}{m_3^2}}+e^{i\beta_{2}}\sqrt{\frac{|m_e|}{2m_\mu}}\right|, \nonumber \\
\left|(U_{MNS})_{21}\right|  & \simeq \frac{1}{\sqrt{2}}\sqrt{\frac{|m_1|}{m_2+|m_1|}} , &
\left|(U_{MNS})_{22}\right|  & \simeq \frac{1}{\sqrt{2}}\sqrt{\frac{m_2}{m_2+|m_1|}} , \nonumber \\ 
\left|(U_{MNS})_{23}\right|  & \simeq \frac{1}{\sqrt{2}},\nonumber \\
\left|(U_{MNS})_{31}\right|  & \simeq \frac{1}{\sqrt{2}}\sqrt{\frac{|m_1|}{m_2+|m_1|}} , & 
\left|(U_{MNS})_{32}\right|  & \simeq \frac{1}{\sqrt{2}}\sqrt{\frac{m_2}{m_2+|m_1|}} , \nonumber \\
\left|(U_{MNS})_{33}\right|  & \simeq \frac{1}{\sqrt{2}}. \label{abs-u}
\end{align}
\par
From Eqs.~(\ref{stdrepLep}) and (\ref{abs-u}), 
the neutrino oscillation angles and phases of the model are 
related to the lepton masses as follows: 
\begin{eqnarray}
\tan^2\theta_{\mbox{{\tiny solar}}}& =&\frac{|(U_{MNS})_{12}|^2}{|(U_{MNS})_{11}|^2}\simeq \frac{|m_1|}{m_2}\ ,\label{eq30300}\\
\sin^2 2\theta_{\mbox{{\tiny atm}}}& =&4|(U_{MNS})_{23}|^2|(U_{MNS})_{33}|^2\simeq 1\ ,
\label{eq30310} \\
|(U_{MNS})_{13}|^2 &\simeq& \left|\sqrt{\frac{|m_1|m_2}{m_3^2}}+e^{i\beta_{2}}\sqrt{\frac{|m_e|}{2m_\mu}}\right|^2. \label{eq30320} \\
\delta_\nu &\simeq& -\mbox{arg}
\left(
\sqrt{\frac{|m_1|m_2}{m_3^2}}+e^{i\beta_{2}}\sqrt{\frac{|m_e|}{2m_\mu}}
\right), \\
\phi_2 &\simeq& \phi_3 \simeq -\frac{\pi}{2}.
\end{eqnarray}
It should be noted that the present model leads to the same results for $\theta_{\mbox{{\tiny solar}}}$ and $\theta_{\mbox{{\tiny atm}}}$ as the model in Ref.~\cite{Matsuda2}, 
while a different feature for $|(U_{MNS})_{13}|^2$ is derived.
\par
On the other hand, we have~\cite{Garcia} a experimental bound for $|(U_{MNS})_{13}|_{\mbox{\tiny exp}}^2$
from the CHOOZ~\cite{chooz}, solar~\cite{sno}, and atmospheric neutrino 
experiments~\cite{skamioka}. 
From the global analysis of the SNO solar neutrino experiment~\cite{sno,Garcia}, 
we have $\Delta m_{12}^2$ and $\tan^2 \theta_{12}$ for the large mixing angle (LMA) Mikheyev-Smirnov-Wolfenstein (MSW) solution~\cite{MSW}.
From the atmospheric neutrino experiment~\cite{skamioka,Garcia} , 
we also have $\Delta m_{23}^2$ and $\tan^2 \theta_{23}$. 
These experimental data with $3\sigma$ range are given by 
\begin{eqnarray}
& &|U_{13}|_{\mbox{\tiny exp}}^2 <  0.054 \ \label{mat20820} \ ,\\
& &\Delta m_{12}^2=m_2^2-m_1^2= \Delta m_{\mbox{{\tiny sol}}}^2
=(5.2-9.8) \times 10^{-5}\, \mbox{eV}^2, \label{mat20830} \\
& &\tan^2 \theta_{12}=\tan^2 \theta_{\mbox{{\tiny sol}}}=0.29-0.64 \ ,\label{eq20501}\\
& &\Delta m_{23}^2=m_3^2-m_2^2 \simeq \Delta m_{\mbox{{\tiny atm}}}^2
= (1.4-3.4) \times 10^{-3}\, \mbox{eV}^2, \label{mat20831}\\ 
& &\tan^2 \theta_{23} \simeq \tan^2 \theta_{\mbox{{\tiny atm}}}=0.49-2.2 \ . \label{mat208302}
\end{eqnarray}
Hereafter, for simplicity, we take  $\tan^2 \theta_{\mbox{{\tiny atm}}} \simeq 1$.
Thus, by combining the present model with the mixing angle \(\theta_{\mbox{{\tiny sol}}}\),
we have 
\begin{equation}
\frac{m_1}{m_2} \simeq \tan^2\theta_{\mbox{{\tiny sol}}}=0.29 - 0.64. 
\label{ratio}
\end{equation}
Therefore we predict the neutrino masses as follows.
\begin{eqnarray}
m_1^2 & = & (0.48-6.8) \times 10^{-5} \  {\rm eV^2} \ ,\nonumber \\
m_2^2 & = & (5.7-16.6) \times 10^{-5} \  {\rm eV^2} \ ,  \label{neu-mass}\\
m_3^2 & = & (1.5-3.6) \times 10^{-3} \  {\rm eV^2} \ .\nonumber
\end{eqnarray}
\par
Let us mention a specific feature of the model. Our model predicts a rather large value for  \(|(U_{MNS})_{13}|\) as
\begin{equation}
|(U_{MNS})_{13}|^2 \simeq \left|\sqrt{\frac{|m_1|m_2}{m_3^2}}+e^{i\beta_{2}}\sqrt{\frac{|m_e|}{2m_\mu}}\right|^2\simeq \frac{|m_1|m_2}{m_3^2}=(0.041 - 9.6)\times10^{-2} \ . \label{mat20870} 
\end{equation}
The predicted value for $|(U_{MNS})_{13}|$ in Eq.~(\ref{mat20870}) is close to  the present experimental constraints 
Eq.~(\ref{mat20820}) in contrast to previously proposed model~\cite{Koide}\cite{Matsuda2}. Therefore our model will  be checked in neutrino factories in near future.
\par
In the preset model, the neutrino mass matrix \(M_\nu\) is given by 
\begin{equation}
M_\nu \simeq P^\dagger_\nu 
\left(
\begin{array}{ccc}
0                & \sqrt{2|m_1|m_2} & 0 \\
\sqrt{2|m_1|m_2} & m_3/2            & m_3/2 \\
0                & m_3/2            & m_3/2 
\end{array}
\right)P^{*}_\nu.\label{nu-matrix}
\end{equation}
Now we discuss the requirements for the mass matrix elements of  $M_D$ and $M_R$ to realize the above structure for 
$M_\nu$. In our model we have assumed the seesaw mechanism $M_\nu =-M_D M_R^{-1} M_D^T$ and following structure for $M_D$ and $M_R$.
\begin{equation}
M_D = P^\dagger_D
\left(
\begin{array}{ccc}
0   & a_D   & 0 \\
a_D & b_D   & c_D \\
0   & c_D & d_D 
\end{array}
\right)
P_D, \quad 
M_R =
\left(
\begin{array}{ccc}
0 & a_R & 0 \\
a_R & b_R & c_R \\
0 & c_R & d_R 
\end{array}
\right),
\end{equation}
where $P_D =\mbox{diag}\left(e^{i\alpha_{D1}}, e^{i\alpha_{D2}}, e^{i\alpha_{D3}}\right)$. Here we assume a real symmetric $M_R$ for simplicity. In this case, we have
\begin{eqnarray}
M_\nu 
&= &- M_D M_R^{-1} M_D^T \\
&\simeq &
-P^{\dagger}_{\nu}
\left(
\begin{array}{ccc}
0               & \frac{a_D^2}{a_R}   & 0 \\
\frac{a_D^2}{a_R} & \frac{c_D^2}{d_R} & \frac{c_Dd_D}{d_R} \\
0               & \frac{c_Dd_D}{d_R}      & \frac{d_D^2}{d_R} 
\end{array}
\right) P^{*}_{\nu} \quad \mbox{   (for \(a_D \ll c_D, d_D \))}
\end{eqnarray}
where $P_{\nu} =\mbox{diag}\left(e^{i(\alpha_{D3} - \alpha_{D2})},  e^{-i(\alpha_{D3} - \alpha_{D2})}, 1\right)$. 
Therefore, the following conditions should be satisfied in order to realize our $M_{\nu}$ in Eq.~(\ref{nu-matrix}),
\begin{equation}
\frac{a_D^2}{a_R} \ll \frac{c_D^2}{d_R} \simeq \frac{c_Dd_D}{d_R} \simeq \frac{d_D^2}{d_R}.
\end{equation}
\par
Namely, it turns out that the large lepton mixing angle is realized through the seesaw mechanism 
by using the following $M_D$ and $M_R$,
\begin{align}
M_D &= P^\dagger_D \left(
\begin{array}{ccc}
0   & a_D   & 0 \\
a_D & *   & d_D \\
0   & d_D & d_D 
\end{array}
\right)
P_D , & 
M_R &=
\left(
\begin{array}{ccc}
0 & a_R & 0 \\
a_R & * & * \\
0 & * & d_R 
\end{array}
\right),\label{condition-1}
\end{align}
with $c_D=d_D$ and a hierarchy conditon 
\begin{equation}
\left(\frac{a_D}{d_D}\right)^2 \ll \frac{a_R}{d_R}. \label{condition-2}
\end{equation}
It should be noted that the components $b_D$ in $M_D$, $b_R$, and $c_R$ in $M_R$ which are denoted as asterisks 
are not important for reproducing the large lepton mixing angle at all.
\section{conclusion}
We have reconsidered the mass matrix model  
with a universal and seesaw-invariant form of four-zero structure given by 
\begin{eqnarray}
M_f&=&P^\dagger_f \left(
\begin{array}{ccc}
0   & a_f & 0   \\ 
a_f & b_f & c_f\\ 
0   & c_f & d_f
\end{array}
\right) P_f, \quad \mbox{for } f=u,d,\mbox{and }e \\
M_f&=&P^\dagger_f \left(
\begin{array}{ccc}
0   & a_f & 0   \\ 
a_f & b_f & c_f\\ 
0   & c_f & d_f
\end{array}
\right) P^{*}_f, \quad \mbox{for } f=\nu
\end{eqnarray}
\par
The analytical expressions for the CKM quark mixing maitrix are derived as functions of 
the four parameters $x_u$, $x_d$, $\alpha_2$, and $\alpha_3$.
We do fine tuning of the parameters so as to reproduce the experimental data. 
It turns out that the CKM quark mixing matrix can be consistent 
with the data at the special value of the parameter given by 
$\alpha_2 \simeq \pi/2$ and in the allowed regions among  $\alpha_3$, $x_u$, and $x_d$ 
as shown in Fig.~1, Fig.~2, and Fig.~3.
\par
We have also analyzed the MNS lepton mixing matrix analytically and shown that 
it is consistent with the observed large lepton mixings. 
The model predicts a relatively large (1,3) element for the MNS lepton mixing matrix element:
\begin{equation}
|(U_{MNS})_{13}|^2 \simeq \frac{|m_1|m_2}{m_3^2} \simeq (0.041 - 9.6)\times10^{-2} \, , 
\end{equation} 
which is close to the experimental upper bound at present. 
Therefore a determination of the finite value for $|(U_{MNS})_{13}|^2$ in near future experiment 
will be expectable in our model.
\par
We have assumed the seesaw mechanism $M_\nu =-M_D M_R^{-1} M_D^T$ and the same four-zero structure for $M_D$ and $M_R$.
Within this framework, we have derived the conditions given by Eqs.~(\ref{condition-1}) and (\ref{condition-2}) 
for the components of $M_D$ and $M_R$ to realize our structure for $M_\nu$.


\begin{acknowledgments}
We thank M. Bando and M. Obara for pointing out the work in Ref~\cite{Kim} to the authors.
\end{acknowledgments}



\newpage
\begin{figure}[htbp]
\begin{center}
\includegraphics{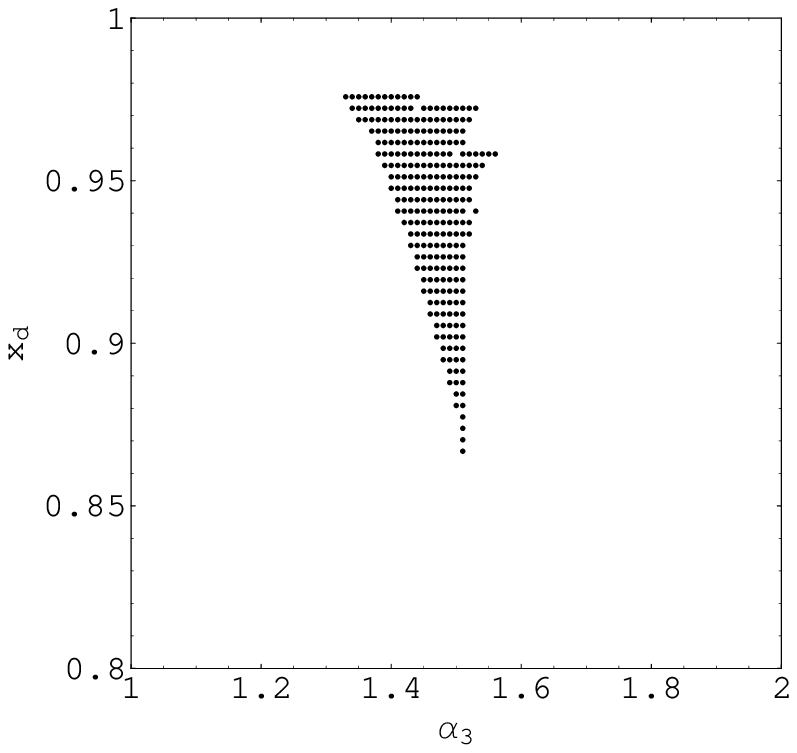}
\end{center}
\caption{%
The allowed region in the $\alpha_3$ - $x_d$ parameter plane.
Dotted regions are allowed from the experimental data for the CKM quark mixing matrix elements. 
}
\label{fig1}
\end{figure}

\begin{figure}[htbp]
\begin{center}
\includegraphics{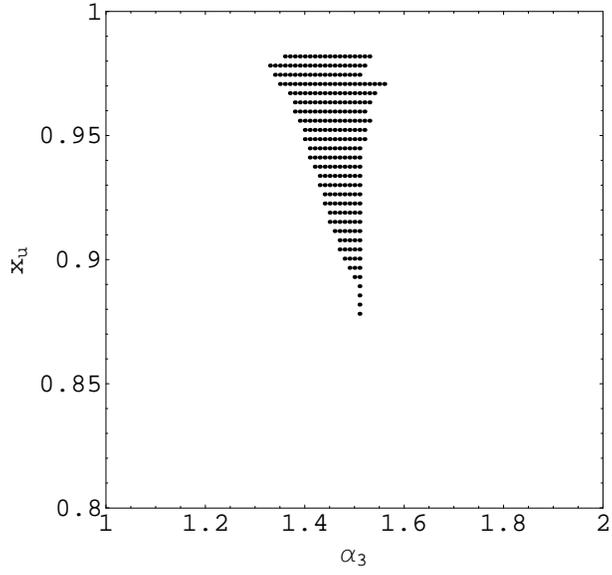}
\end{center}
\caption{%
The allowed region in the $\alpha_3$ - $x_u$ parameter plane. 
}
\label{fig2}
\end{figure}

\begin{figure}[htbp]
\begin{center}
\includegraphics{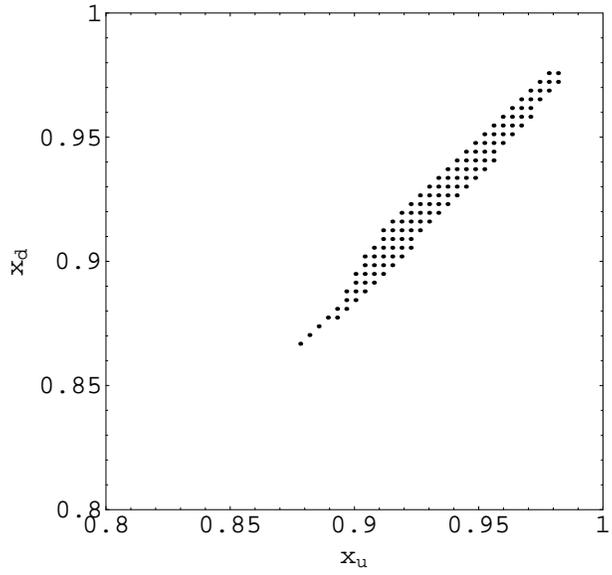}
\end{center}
\caption{%
The allowed region in the $x_u$ - $x_d$ parameter plane. 
}
\label{fig3}
\end{figure}

\end{document}